\documentclass[twocolumn]{aastex631}

\def\msun{{\rm\,M_\odot}}
\usepackage{amsmath} 
\usepackage{multirow}

\begin{document}
\title{Redshift Evolution of the Ratio of Supermassive Black Hole Mass to Stellar Mass}

\author{Ziyong Wu}
\affiliation{Institute of Astronomy, School of Physics, Zhejiang University, Hangzhou 310037, China}
\affiliation{Purple Mountain Observatory, Chinese Academy of Sciences, 10 Yuanhua Road, Nanjing 210033, China}
\affiliation{School of Astronomy and Space Sciences, University of Science and Technology of China, Hefei 230026, China}

\author{Renyue Cen}
\altaffiliation{renyuecen@zju.edu.cn}
\email{renyuecen@zju.edu.cn}
\affiliation{Institute of Astronomy, School of Physics, Zhejiang University, Hangzhou 310037, China}
\affiliation{Center for Cosmology and Computational Astrophysics, Institute for Advanced Study in Physics, Zhejiang University, Hangzhou, China}

\author{Romain Teyssier}
\affiliation{Department of Astrophysical Sciences, Princeton University, Peyton Hall, r inceton, NJ 08544, USA}





\begin{abstract}
We run and analyze a suite of high-redshift zoom-in cosmological simulations with varying supernova feedback and supermassive black hole (SMBH) accretion prescriptions to study the joint evolution of stellar and SMBH mass in high-redshift galaxies down to $z=10$. The simulations reproduce the observed high-$z$ $M_{\mathrm{BH}}/M_{\star}$ relation if super-Eddington accretion is allowed prior to the final self-regulated phase. To extend the evolution to lower redshift, we model subsequent black hole and host growth using analytic halo assembly histories combined with a redshift-dependent effective Eddington duty cycle, $f_{\rm duty}=0.0004(1+z)^3$, calibrated to observations at $z\le 6$, with a conservative uncertainty range at higher redshift. Within this framework, we find that $M_{\mathrm{BH}}/M_{\star}$ exhibits a broad peak at $z\sim 7$--10, reaching values from a few percent up to $\sim 30\%$, followed by a steady, approximately power-law decline toward $z=0$. The model predicts $M_{\mathrm{BH}}/M_{\star} \sim (0.002,0.003,0.006,0.016,0.071,0.156)$ at $z=(0,1,2,3,5,10)$, in good agreement with available observations. 
This evolution is driven by rapid SMBH growth at high redshift, with effective mass e-folding times shorter than those of stellar mass, while at later times galaxy growth dominates, leading to the decline in $M_{\mathrm{BH}}/M_{\star}$. These results demonstrate that the emergence of a high-redshift peak and subsequent decline is a robust outcome despite uncertainties in the duty-cycle normalization.
\end{abstract}


\section{Introduction} \label{sec:intro}

Supermassive black holes (SMBHs) with masses of $10^{7}$--$10^{10}\,M_\odot$ have been observed in galaxies less than a billion years after the Big Bang \citep[e.g.,][]{2011Mortlock, 2015Wu,2018Banados}, posing a significant challenge to models of early structure formation. A key question is how these massive black holes assembled so rapidly and how their growth relates to that of their host galaxies. At low redshift, observations reveal a tight correlation between SMBH mass and host stellar or bulge mass \citep{2013ARA&A..51..511K,2013ApJ...764..184M,2015ApJ...813...82R,2020ARA&A..58..257G}, but whether this relation holds---or even applies---in the early Universe remains uncertain.

Prior to \textit{JWST}, evidence for overmassive SMBHs at high redshift had already emerged from optical, near-infrared, and sub-millimeter observations. Large quasar surveys such as SDSS and UKIDSS uncovered luminous quasars at $z\gtrsim 6$ hosting black holes with $M_{\mathrm{BH}}\sim 10^{9-10}\,M_\odot$, challenging the short timescales available for early black hole growth \citep{2011Mortlock,2015Wu,2023ARA&A..61..373F}. ALMA measurements of [C\,\textsc{ii}] and CO emission further revealed that many of these quasars reside in comparatively modest host galaxies, implying BH-to-galaxy mass ratios significantly elevated relative to the local $M_{\rm BH}$--$M_\star$ relation \citep{2018ApJ...854...97D,2013ApJ...773...44W,2022A&A...668A.121S}.

Recent \textit{JWST} observations have dramatically extended these studies to lower stellar masses and higher redshifts. A striking and increasingly robust picture has emerged: nearly all galaxies observed at $z>3$ with $M_\star \lesssim 10^9\,M_\odot$ appear to host SMBHs that are overmassive by 1--2 dex relative to the local relation \citep{2023A&A...677A.145U,2024MNRAS.533.4287U,2023ApJ...959...39H,2023ApJ...957L...7K,2023Natur.619..716C,2023ApJ...957L...3P,2024A&A...691A.145M,2024Natur.627...59M,2024Natur.628...57F,2024Natur.636..594J,2024ApJ...960L...1N,2025arXiv251007376J}. Whether this discrepancy reflects genuine evolution or observational selection effects remains debated \citep{2007ApJ...670..249L,2010ApJ...713...41S,2023ApJ...954..173L}. It may also stem from SMBH mass estimates based on methods calibrated only at low redshift \citep{2024ApJ...969L..18A}. Moreover, different scaling relations for spirals and ellipticals imply that adopting the spiral relation at high redshift can reduce the tension \citep{2018ApJ...869..113D,2025arXiv250310958G}.

In this work, we adopt a hybrid approach to investigate the evolution of the SMBH-to-stellar mass ratio across cosmic time. We combine (1) a suite of high-resolution zoom-in simulations of massive galaxies at $z>6$, which follow the early growth of black holes under different supernova (SN) feedback strengths and accretion prescriptions, including models that allow super-Eddington accretion (i.e., accretion rates exceeding the classical Eddington limit, $\dot{M}_{\rm acc}>\dot{M}_{\rm Edd}$), with (2) an analytic framework that connects these early phases to lower redshift using empirically calibrated relations between halo mass, stellar mass, and SMBH accretion activity. The analytic model follows halo mass assembly and incorporates a redshift-dependent effective AGN duty cycle, $f_{\rm duty} = 0.0004(1+z)^3$, motivated by observational constraints at $z=0$--6. This combined framework enables us to trace the coevolution of black holes and their host galaxies from early times to the present day.

\section{METHODOLOGY AND
SIMULATION DETAILS} \label{sec:method}

We perform a cosmological hydrodynamic simulation using the adaptive mesh refinement (AMR) code, RAMSES \citep{2002A&A...385..337T}. The initial conditions are generated with the MUSIC software \citep{2011MNRAS.415.2101H}, adopting cosmological parameters ($\Omega_m$ = 0.288, $\Omega_\Lambda$ = 0.712, $\Omega_b$ = 0.045, $H_0$ = 69.33 $\mathrm{km}s^{-1}\mathrm{Mpc}^{-1}$, $n_s$ = 0.971, and $\sigma_8$ = 0.830), consistent with the WMAP9 results \citep{2013ApJS..208...19H}. The simulation box, with a comoving volume of $(6\, \rm{Mpc})^3$, is initialized with $128^3$ root cells. High-resolution dark matter particles with a mass of $2448 \msun$ are used for the zoom-in region of $(0.52\, \rm{Mpc})^3$ (comoving), which includes two halos with $2 \times 10^9 \msun < M_{halo} < 5 \times 10^9 \msun$ at $z \sim 10$. The Lagrangian volume, or mask, is defined by a scalar quantity that is advected passively with the flow throughout the simulation. Initially, the passive scalar takes a value of 1 inside the mask and 0 outside. Refinement is allowed in regions where the passive scalar exceeds $10^{-3}$, provided that at least one of the following conditions is satisfied: i) the number of dark matter particles within the cell exceeds 8, or ii) the total baryonic mass in the cell exceeds $3060 \msun$. These criteria enable the simulation to achieve a maximum spatial resolution of $\Delta x_{\text{min}} \approx 17.9 \mathrm{pc}$ (physical) at redshift $z = 10$.

A detailed description of the simulation setup is provided in \citet{wu2025fastsupermassiveblackholes}. Here, we highlight the key aspects:

\begin{enumerate}
    \item \textbf{Star formation.} We adopt a star formation number density threshold of $n_0 = 65\,\mathrm{H\ cm}^{-3}$ and a star formation efficiency of $\epsilon_\star = 0.1$, following the Schmidt law,
    \begin{equation}
        \dot\rho_\star = \epsilon_\star \frac{\rho}{t_{\rm ff}},
    \end{equation}
    where $\dot\rho_\star$ is the star formation rate (SFR) density, and $t_{\rm ff} = \sqrt{3\pi / (32 G \rho_{\rm gas})}$ is the local free-fall time of the gas \citep{2006A&A...445....1R, 2008A&A...477...79D}.

    \item \textbf{Black hole seeding.} BH seeds are formed using the sink particle algorithm \citep{1995MNRAS.277..362B, 2004ApJ...611..399K, 2014MNRAS.445.4015B, 2017MNRAS.469..295B} when all of the following criteria are met:
    \begin{enumerate}
        \item the halo mass exceeds $4\times10^4\,M_\odot$;
        \item the gas clump mass exceeds $10^4\,M_\odot$;
        \item the average density within a four-cell sphere exceeds the star formation threshold
        \item the peak gas density is more than three times the star formation threshold.
    \end{enumerate}
    These criteria identify dense, gravitationally collapsing gas clumps where black hole formation is expected. Each seed is initialized with a mass of $M_{\rm seed}=10^3\,M_\odot$, corresponding to light black hole seeds, comparable in mass to remnants of Population III stars.

    \item \textbf{SMBH accretion and feedback.} SMBH accretion follows the Bondi--Hoyle--Lyttleton prescription \citep{1939PCPS...35..405H, 1944MNRAS.104..273B, 1952MNRAS.112..195B}, with AGN feedback operating in two modes: a \textit{kinetic} mode, active at low accretion rates and modeled as a jet-like outflow, and a \textit{thermal} mode, which dominates at high but sub-Eddington accretion rates and heats the surrounding gas via thermal energy injection \citep{2012MNRAS.420.2662D}. The transition between the two regimes is determined by the Eddington ratio, $\lambda_{\rm Edd} \equiv \dot{M}_{\rm acc}/\dot{M}_{\rm Edd}$. Since a constant radiative efficiency is assumed in our model, this definition is equivalent to the luminosity-based form $\lambda_{\rm Edd}=L/L_{\rm Edd}$.
\end{enumerate}

We conduct eight zoom-in cosmological simulations with identical initial conditions but different feedback prescriptions. In all simulations, accretion is permitted to exceed the Eddington rate, with a cap set at three times the Eddington limit. For clarity, each simulation is assigned a descriptive label: runs without feedback are denoted ``No feedback"; runs including only supernova (SN) feedback are denoted ``Only SN"; runs including only AGN feedback are denoted ``Only AGN"; and runs including both SN and AGN feedback are labeled according to the specific SN feedback model adopted. Additionally, we include a simulation with an enhanced SN energy in the kinematic feedback model, effectively representing hypernova feedback. 

The nature of black hole accretion at high redshift remains uncertain, particularly regarding the prevalence of sustained super-Eddington growth. To assess the sensitivity of the evolution of $M_{\rm BH}/M_\star$ to this uncertainty, we compare this suite of super-Eddington simulations that span multiple SN feedback prescriptions, with an otherwise identical run in which accretion is capped at the Eddington limit for a representative feedback model.

The halo mass lies in the range $M_{\rm{halo}} = 2 \times 10^9 \, M_\odot$ to $5 \times 10^9 \, M_\odot$. The dark matter distribution traces the underlying large-scale potential, while the gas shows filamentary structures feeding into the halo center. The impact of SN feedback is most evident in the stellar component (bottom row). In the mechanical feedback case, numerous small stellar clumps are present, reflecting the efficiency of this model in regulating but not entirely suppressing star formation. In contrast, the delayed cooling feedback model strongly suppresses star formation, leading to a much smoother stellar distribution with fewer visible clumps. These differences highlight the sensitivity of early stellar assembly to the adopted SN feedback model.

In addition to our two fiducial halos, we have performed two additional zoom-in simulations targeting halos in distinct large-scale environments, including a void and a cluster region. All simulations are run with identical numerical resolution and physical prescriptions with an enhanced SN energy in the kinematic feedback model, differing only in their large-scale environment and assembly history.

\begin{figure*}[htbp]
    \center
    \includegraphics[width=\textwidth]{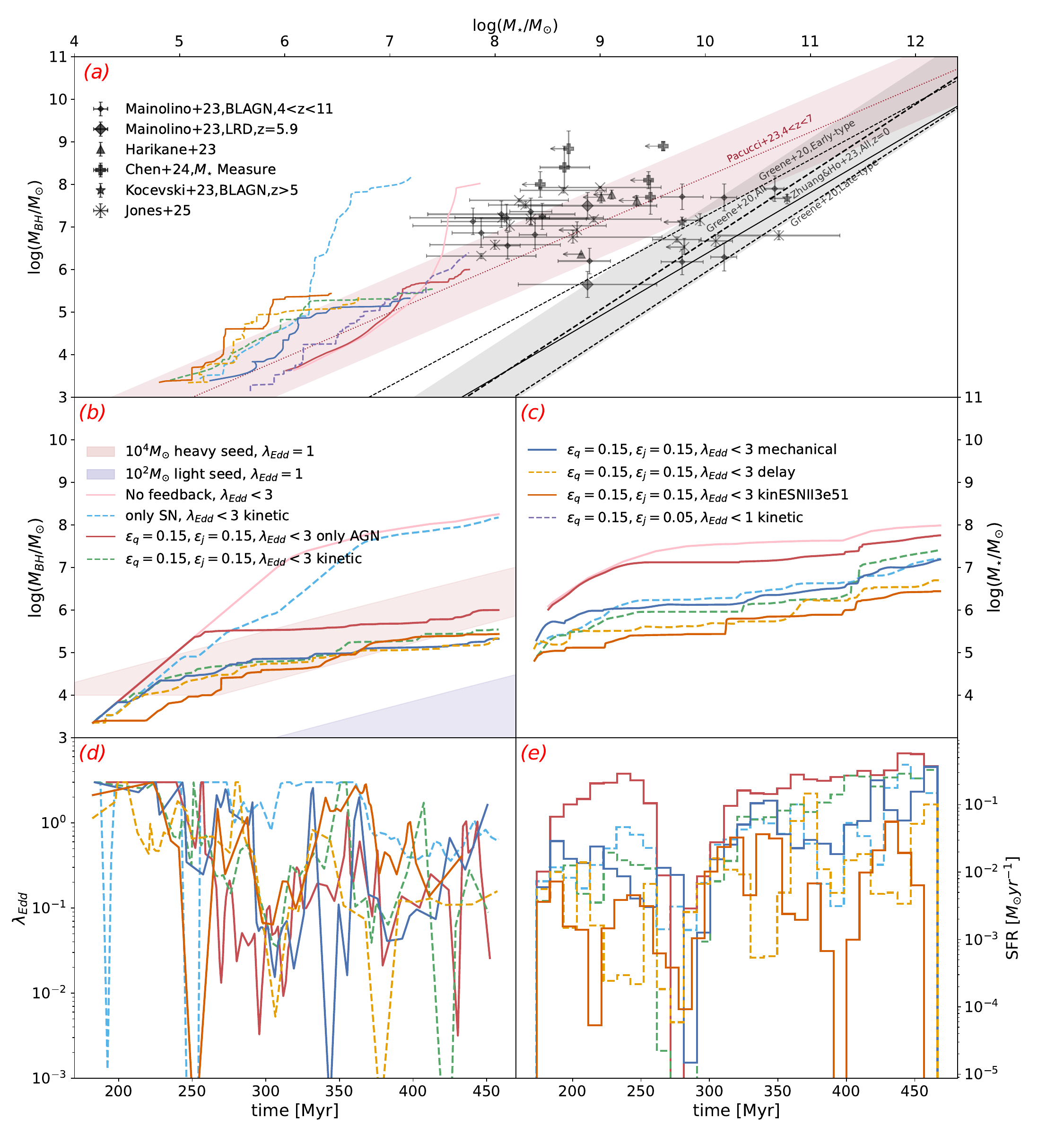}\\
    \caption{
(a) The $M_{\rm BH}$–$M_\star$ relation combining our simulations with observational constraints and empirical scaling relations. High-redshift broad-line AGNs, including so-called ``little red dots'' (LRDs) from \citet{2024A&A...691A.145M}, \citet{2023ApJ...954L...4K}, \citet{2023ApJ...959...39H}, and \citet{2025arXiv250723066S}, are shown together with six LRDs with stellar mass measurements from \citet{2025ApJ...983...60C}. The black dashed curves indicate the local relations for all, early-type, and late-type galaxies \citep{2020ARA&A..58..257G}. The solid line denotes the AGN scaling relation evolved to $z=0$ \citep{2023NatAs...7.1376Z}, while the dotted curve shows the relation at $4<z<7$ \citep{2023ApJ...957L...3P}; shaded bands represent the $1\sigma$ intrinsic scatter. Our simulations are overplotted as evolutionary tracks in different colors (see legend): no-feedback (red solid), SN-kinetic (yellow dotted), AGN-only (purple solid), AGN+SN-kinetic (blue dotted), AGN+SN-mechanical (blue solid), AGN+SN-delayed cooling (orange solid), and AGN+SN-kinetic with $3\times10^{51}\,\mathrm{erg}$ (green solid).  
(b) Black hole mass growth, compared to the Eddington-limited predictions for $10^2\,M_\odot$ (blue shaded) and $10^4\,M_\odot$ (red shaded) seed black holes.  
(c) Stellar mass growth.  
(d) Eddington ratios, $\lambda_{\rm Edd}$.  
(e) Star formation rates.  
Each feedback model uses the color and line style shown in the legend.}

    \label{fig:data}
\end{figure*}

\section{SN feedback} \label{sec:feedback}

Supernova feedback is performed with single and instantaneous injections of the cumulative SN energy per stellar population particle. Each stellar particle has an energy and mass
injection budget of
\begin{equation}
\begin{aligned}
E_{\text{SN}} & =10^{51} \mathrm{erg} \quad \eta_{\text{SN}} \frac{m_*}{m_{\text{SN}}} \\
m_{ej} & =\eta_{\text{SN}} m_{\star}
\end{aligned}
\label{eq:SNf}
\end{equation}
respectively, where $\eta_{\text{SN}}$ is the fraction of stellar mass that is recycled into SN ejecta, $m_{\text{SN}}$ is the average stellar mass of a type II SN progenitor, and $m_{\star}$ is the mass of the stellar particle.

We assume that 21\% of the stellar mass is returned to the interstellar medium, of which 7.5\% consists of newly synthesized metals, corresponding to a mass-loading factor of $\eta_{\rm SN} = 0.21$. These values are adopted following previous \textsc{RAMSES}-based studies \citep[e.g.,][]{2012MNRAS.420.2662D,2013MNRAS.429.3068T}.

We use the kinetic feedback model presented in \citet{2008A&A...477...79D}. Here, the trick to overcoming numerical overcooling is to skip the unresolved Sedov-Taylor phase and directly inject the expected collective result of that phase for a stellar population, which is an expanding momentum conserving shock wave (or snowplow). Note, however, that
the injected kinetic energy may subsequently be converted
into thermal energy if shocks develop. SN mass and momentum is injected into gas within a bubble radius of the exploding stellar particle. The free parameters for the method are $f_k$, the fraction of $E_{\text{SN}}$ which is released in kinetic form, $r_{\text{bubble}}$,  the radius of the bubble, and $\eta_{\text{W}}$, the sub-resolution mass loading factor of the
Sedov-Taylor phase, describing how much mass, relative to
the stellar mass, is redistributed from the cell at the bubble
centre to the bubble outskirts.

The redistributed mass consists of two components: one
is the SN ejecta, $m_{\mathrm{ej}}=\eta_{\mathrm{SN}} m_*$, removed from the stellar
particle, the other is the swept up mass,  $m_{\mathrm{sw}}=\eta_{\mathrm{W}} m_*$, removed from the central host cell (no more than 25\% of the
central cell mass is removed, hence for individual feedback
events at relatively low densities it may happen that $m_{\text{sw}}$ is smaller than $\eta_{\mathrm{W}} m_*$). The total wind mass is thus $m_{\mathrm{sw}}=\eta_{\mathrm{W}} m_*$, which is redistributed uniformly (i.e. uniform
density) to all cells inside the bubble.

The kinetic energy, $f_k E_{\text{SN}}$, is likewise distributed to the
bubble cells, but with an injected velocity (directed radially
away from the stellar particle) that increases linearly with
distance from the centre, such as to approximate the ideal
Sedov-Taylor solution:

\begin{equation}
\mathbf{v}\left(\Delta m_{\text {cell }}\right)=f_{\mathrm{N}} v_{\mathrm{W}} \frac{\mathbf{r}_{\text {cell }}}{r_{\text {bubble }}}
\end{equation}

where $\Delta m_{\text {cell }}$ is the mass added to the cell, $r_{\text{cell}}$ is the position of the centre of the cell relative to the stellar particle, $f_{N} \sim 1$  is a bubble normalisation constant required to ensure that the total redistributed energy is equal to $f_k E_{\text{SN}}$, and 
\begin{equation}
v_{\mathrm{W}}=\sqrt{\frac{2 f_{\mathrm{k}} E_{\mathrm{SN}}}{m_{\mathrm{W}}}}=3,162 \mathrm{~km} / \mathrm{s} \sqrt{\frac{f_{\mathrm{k}}}{1+\eta_{\mathrm{W}} / \eta_{\mathrm{SN}}}}
\end{equation}
is the unnormalised wind velocity, where we used Eq.\ref{eq:SNf} for the latter equality. Note that this is the velocity of the
added mass, i.e. each cell gains momentum
\begin{equation}
\Delta \mathbf{p}=\mathbf{v}\left(\Delta m_{\text {cell }}\right) \Delta m_{\text {cell }} \propto \sqrt{f_{\mathrm{k}} \eta_{\mathrm{SN}}\left(\eta_{\mathrm{SN}}+\eta_{\mathrm{W}}\right)}
\end{equation}
so if the mass already in the cell is substantial compared to
the added mass, the resulting velocity change can be small.
The injection is performed in the mass-weighted frame of
the SN particle (with $m_{\text{ej}}$ and host cell (with $m_{\text{sw}}$. The
remaining thermal energy,  $(1 - f_k) E_{\text{SN}}$, is then distributed uniformly between the bubble cells.

In this work, we use fiducial parameters $f_k = 0.5, \eta_{\text{W}} = 5$ and $r_{\text{bubble}}=35\,\mathrm{pc}$. These values give a velocity for the gas ejected from the
central cell of $v_{\text{W}} \approx 140\, \mathrm{km/s}$. Our choice of $f_k = 1$ implies that there have been neither radiative losses nor momentum cancellation from the set of unresolved individual SNe inside the bubble.

This prescription serves as our fiducial SN feedback model. Based on this scheme, we also run a simulation adopting a larger single SN energy of $3 \times 10^{51}\,\mathrm{erg}$.

We also perform a simulation based on the mechanical feedback scheme of \citet{2014ApJ...788..121K}, in which SN energy and momentum are injected according to the evolutionary stage of the Sedov–Taylor blast wave. This method captures the correct momentum and energy transfer across all phases of the blastwave evolution and accounts for the continuous distribution of massive star lifetimes, spanning 3–40~Myr \citep{2015MNRAS.451.2900K}.

In addition, we test a simulation employing the delayed-cooling model of \citet{2013MNRAS.429.3068T}, where radiative cooling is temporarily suppressed in SN-heated gas. The key free parameter in this model is the dissipation timescale, $t_{\rm delay}$, which governs the duration of turbulent energy retention. Following previous studies \citep{2013MNRAS.429.3068T,2014MNRAS.444.2837R,2015MNRAS.447.1353M,2016MNRAS.457.1722R}, we adopt a fiducial value of $t_{\rm delay} = 20\,\mathrm{Myr}$.

\begin{figure*}[htbp]
    \center
    \includegraphics[width=\textwidth]{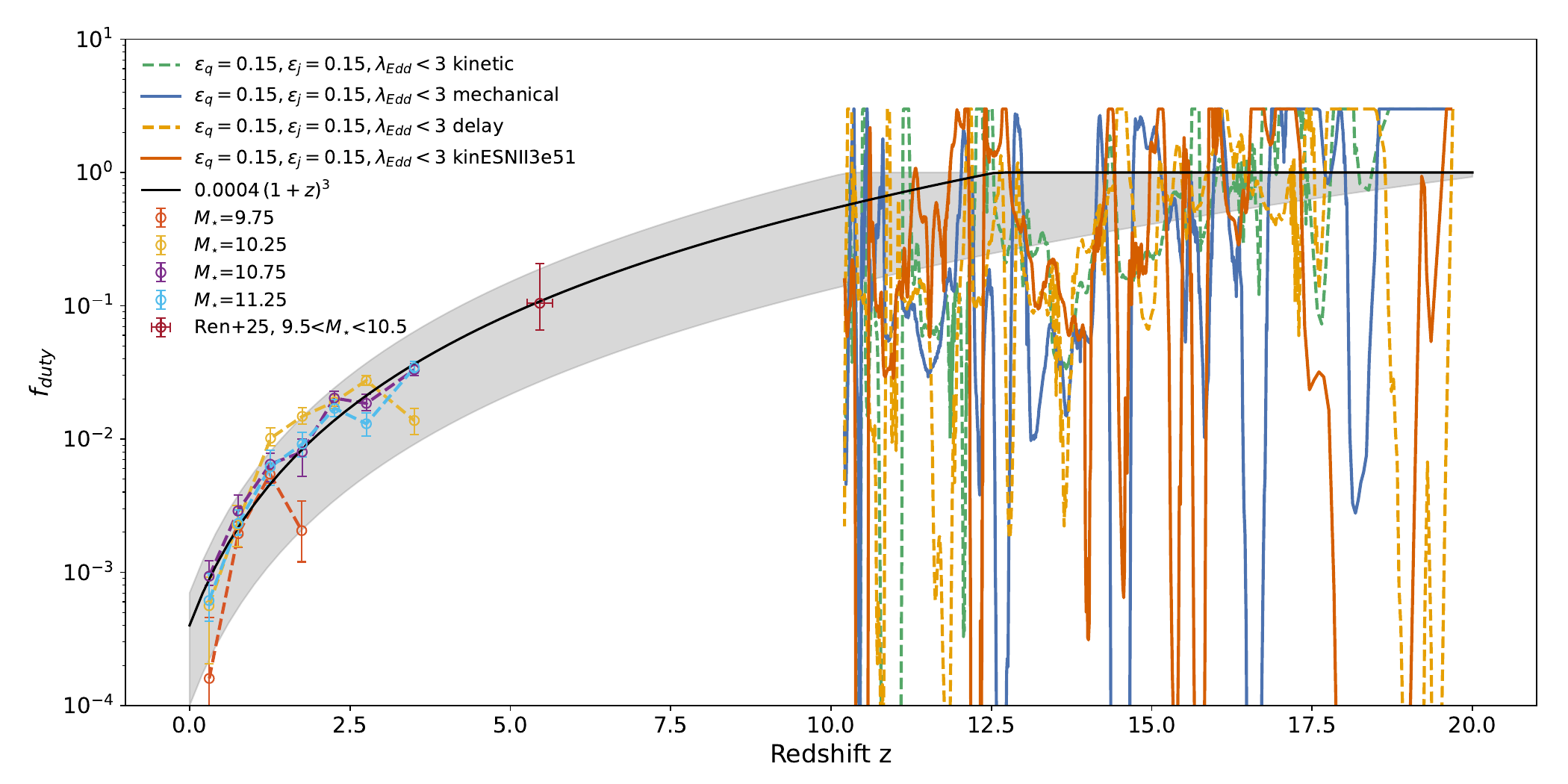} \\
    \caption{The duty cycle in our model, $f_{\rm duty}(z) = 0.0004\,(1+z)^3$ (black solid line), is shown together with a grey shaded region indicating the adopted uncertainty in the duty-cycle normalization arising from observational calibration and extrapolation beyond $z\lesssim 6$. Colored points indicate measurements for galaxies of different stellar masses (as specified in the legend), with $1\,\sigma$ confidence intervals. These observational duty cycles are inferred by multiplying the AGN fraction by the mean accretion rate of the full galaxy population, following the methodology of \citet{2017MNRAS.465.3390A} and \citet{2025MNRAS.544..211R}. Colored curves show the duty cycles measured directly from our simulations with different feedback prescriptions (legend labels).}
    \label{fig:calibrate}
\end{figure*}

\section{Result} \label{sec:result}
\subsection{Overmassive SMBHs in high redshift}
\label{sec:overmass}

Based on our simulations, we reproduce an $M_{\rm BH}$–$M_\star$ relation that is broadly consistent with the high-redshift relation observed by JWST. In Fig.~\ref{fig:data} (a), following the presentation of \citet{2025arXiv250504609T}, we show the black hole–stellar mass relation at high redshift (red shaded region, $4 < z < 7$) and at low redshift (grey line). A clear offset is apparent: high-$z$ broad-line AGNs are found to host systematically over-massive black holes relative to the local $M_{\rm BH}$–$M_\star$ relation \citep{2024A&A...691A.145M,2024Natur.627...59M,2023ApJ...959...39H,2023ApJ...957L...7K,2024Natur.628...57F,2023ApJ...957L...3P,2024ApJ...966..176Y,2023ApJ...953..180S,2025ApJ...983...60C,2025MNRAS.544..211R,2025arXiv250723066S}. Although there remains considerable debate over whether this relation truly evolves across cosmic time or whether the observed offset is primarily a selection effect \citep{2022ApJ...931L..11L,2025ApJ...981...19L,2025arXiv250622147G,2025arXiv250723066S}, our simulation results predominantly align with the high-redshift relation.

While variations in the SN feedback prescription significantly affect stellar mass growth, the black hole mass follows a remarkably similar evolutionary track across different models. To further explore the impact of energetic feedback, we perform an additional simulation that adopts the same SN feedback model but increases the energy per explosion to $3\times10^{51}\,\mathrm{erg}$, effectively mimicking the contribution of hypernovae. In this high-energy case (green solid line), the black hole mass at $z=10$ is nearly identical to that of the lower-energy case (blue dotted line), yet the stellar mass is suppressed by almost an order of magnitude. This striking contrast demonstrates that, within typical high-redshift halos, black hole growth is largely insensitive to supernova feedback, whereas stellar mass assembly is strongly regulated by it.

The differing responses of black holes and stars to SN feedback naturally explain the elevated black hole–to–stellar mass ratios ($M_{\rm BH}/M_\star$) observed in the JWST era. More efficient SN feedback at high redshift—arising either from intrinsically more energetic explosions per unit stellar mass or from the shallower gravitational potential wells of early galaxies—can drive strong outflows that preferentially suppress star formation without significantly hindering black hole growth. Such a mechanism offers a compelling pathway to reconcile the apparent overmassive black holes observed at early cosmic times with theoretical models of galaxy–SMBH co-evolution.

Here we also estimate the relative growth rates of the black hole and stellar components in our simulation with analytical form.

For the stellar mass, we assume
\begin{equation}
M_{\star} = \int \mathrm{SFR}(t)\, dt ,
\end{equation}
The star formation rate (SFR) follows the empirical relation between halo accretion rate from \citet{2010MNRAS.406.2267F}:
\begin{equation}
\begin{aligned}
\dot{M}_{\rm halo} = {}& 46.1\,M_{\odot}\,{\rm yr}^{-1}
\left(\frac{M_{\rm halo}}{10^{12} M_{\odot}}\right)^{1.1} \\
& \times (1 + 1.11z)\,
\left[\Omega_m(1+z)^3 + \Omega_\Lambda \right]^{1/2},
\end{aligned}
\end{equation}  
\begin{equation}
\mathrm{SFR}(M_{\rm halo}, z) = \epsilon_{\star}\, \frac{\Omega_b}{\Omega_m}\, \dot{M}_{\rm halo}(z).
\end{equation}


For the black hole mass growth, we write
\begin{equation}
M_{\rm BH}(t) = \int \dot{M}_{\rm BH}(t)\, dt,
\end{equation}
where the accretion rate follows an Eddington–ratio and duty–cycle prescription,
\begin{equation}
\dot{M}_{\rm BH}(t) = f_{\rm duty}(t)\, \frac{M_{\rm BH}(t)}{t_{\rm Salp}},
\end{equation}
with $t_{\rm Salp} = 4.5 \times 10^7, (\epsilon_r/0.1),{\rm yr}$ the Salpeter timescale and $\epsilon_r$ the radiative efficiency of the accretion disk. The duty cycle $f_{\rm duty}$ represents the effective fraction of time during which the black hole accretes at the Eddington rate, i.e., the product of the mean Eddington ratio and the fraction of time the AGN is active.

To estimate $f_{\rm duty}$, we follow \citet{2024MNRAS.530.1512A}, who analyzed the redshift evolution of Eddington ratios using the AGN catalog of \citet{2017ApJS..228....9K}. They found that the mean Eddington ratio scales as $\lambda_{\rm Edd} \propto \beta\,\epsilon/(1-\epsilon)\,(1+z)^3$. Motivated by this result, we adopt
\begin{equation}
f_{\rm duty}(z) \propto (1+z)^3.
\end{equation}

The normalization is calibrated using independent observational constraints. Because $f_{\rm duty}$ corresponds to the product of the AGN fraction and the mean accretion rate in the self-regulated regime, we compare our model with measurements from \citet{2017MNRAS.465.3390A} and \citet{2025MNRAS.544..211R} (Fig.~\ref{fig:calibrate}). In their analysis, galaxies hosting black holes with $\lambda_{\rm Edd} > 0.01$ are classified as AGN, allowing both the AGN fraction and the mean Eddington ratio to be estimated in bins of stellar mass. Multiplying these quantities yields the observational estimate of the effective duty cycle shown in Fig.~\ref{fig:calibrate}. Motivated by this comparison, we adopt the redshift-dependent form

\begin{equation}
f_{\rm duty}(z) = 0.0004\,(1+z)^3,
\end{equation}

as the effective Eddington-duty cycle in our model. This prescription is empirically calibrated using AGN observations at $z \le 6$. Its extension to higher redshift is not derived from first principles and should be interpreted as a phenomenological continuation rather than a precise prediction; at $z \gtrsim 6$, this extrapolation carries substantial uncertainty. To account for this, we allow the normalization of the duty-cycle model to vary within a broad, observationally motivated range around the fiducial relation, exploring variations of $\pm 0.3$ dex, consistent with the scatter inferred from AGN constraints at $z \le 6$. We also tested an alternative redshift scaling of the duty cycle, but found it inconsistent with the observational data at $z \le 6$. The range of normalization adopted in our fiducial model is therefore intended to bracket plausible behaviors while remaining fully consistent with current observations.

In Fig.~\ref{fig:calibrate}, we also show the duty cycle measured directly from our simulations. Initially, black holes accrete at high Eddington ratios, reflecting the rapid early growth phase. Around $z \sim 13$, the black holes transition into a self-regulated accretion phase, characterized by a decline in the Eddington ratio (detailed in \citet{wu2025fastsupermassiveblackholes}). Remarkably, this transition in our simulation closely coincides with the redshift of $z \simeq 12.6$ inferred from the observation-calibrated duty cycle model. This agreement suggests that black hole growth naturally transitions from an early phase of rapid, near-Eddington accretion to a later phase of slower, sub-Eddington growth, indicating that our duty cycle prescription captures the key physical processes governing the early evolution of black holes.

The resulting growth ratio between black hole and stellar mass can be expressed as
\begin{equation}
\begin{aligned}
\frac{\Delta M_{\rm{BH}}}{\Delta M_{\star}} & \simeq f_{\rm duty} \frac{\Omega_m}{\Omega_b} \frac{M_{\rm{BH}}}{M^{1.1}_{\rm{halo}}} \frac{t_{\rm{halo}}}{t_{\rm{Salpeter}}} \epsilon_{\star}^{-1} \\
& \simeq 0.002(1+z)^{1/2}\left(\frac{M_{\rm{BH}}}{10^8 \msun}\right)\\
& \left(\frac{M_{\rm{halo}}}{10^{12} \msun}\right)^{-1.1}\left(\frac{\epsilon_{\star}}{0.1}\right)^{-1}
\end{aligned}
\label{eq:eq13}
\end{equation}

This dimensionless ratio quantifies how rapidly the BH grows relative to the stellar component. For comparison, the typical local black hole–to–stellar mass ratio is $M_{\rm BH}/M_\star \simeq 0.002$. Thus, whenever $\Delta M_{\rm BH}/\Delta M_\star > 0.002$, the system would evolve toward the “overmassive” regime. Implementing $M_{\rm BH} = 3\times 10^5 \msun$, $M_{\rm halo} = 4\times 10^9 \msun$ and $\epsilon_{\star} = 0.1$ we attain $\Delta M_{\rm BH}/\Delta M_{\star} = 0.012$ at $z=10$. Therefore, this ratio consistently exceeds 0.002. 

Our result can be compared to the analytic model of \citet{2024ApJ...964..154P}, who derived $M_{\rm BH}/M_\star \propto \xi(z)^{5/6}(1+z)^{5/2}$ from a self-regulated framework. While Eq.~\ref{eq:eq13} does not take the form of a single power-law in redshift, combining it with the virial and self-regulated scalings adopted by \citet{2024ApJ...964..154P} yields a similarly steep effective evolution, of order $(1+z)^3$. Thus, both approaches indicate a strong increase of the black hole-to-stellar growth ratio toward high redshift. The key difference lies in the underlying assumptions: \citet{2024ApJ...964..154P} derive an analytic scaling based on self-regulation tied to halo circular velocity, whereas our model follows the coupled, time-dependent growth of black holes, stars, and their host halos in a cosmological context, incorporating a redshift-dependent duty cycle.

This analytic estimate, and its consistency with previous analytic frameworks, agrees with our numerical results that the simulations consistently produced overmassive black holes while maintaining realistic stellar masses and star formation histories. The persistence of this offset across feedback strengths and accretion prescriptions demonstrates that the high $M_{\rm BH}$–$M_\star$ ratio is a physically plausible consequence of efficient early gas inflows and temporarily super-Eddington accretion, rather than a product of numerical artifacts or parameter fine-tuning. This scaling argument therefore supports the interpretation that the overmassive SMBHs observed at high redshift are a natural outcome of early cosmic conditions that favor rapid BH growth relative to star formation.

\begin{figure*}[htbp]
    \center
    \includegraphics[width=\textwidth]{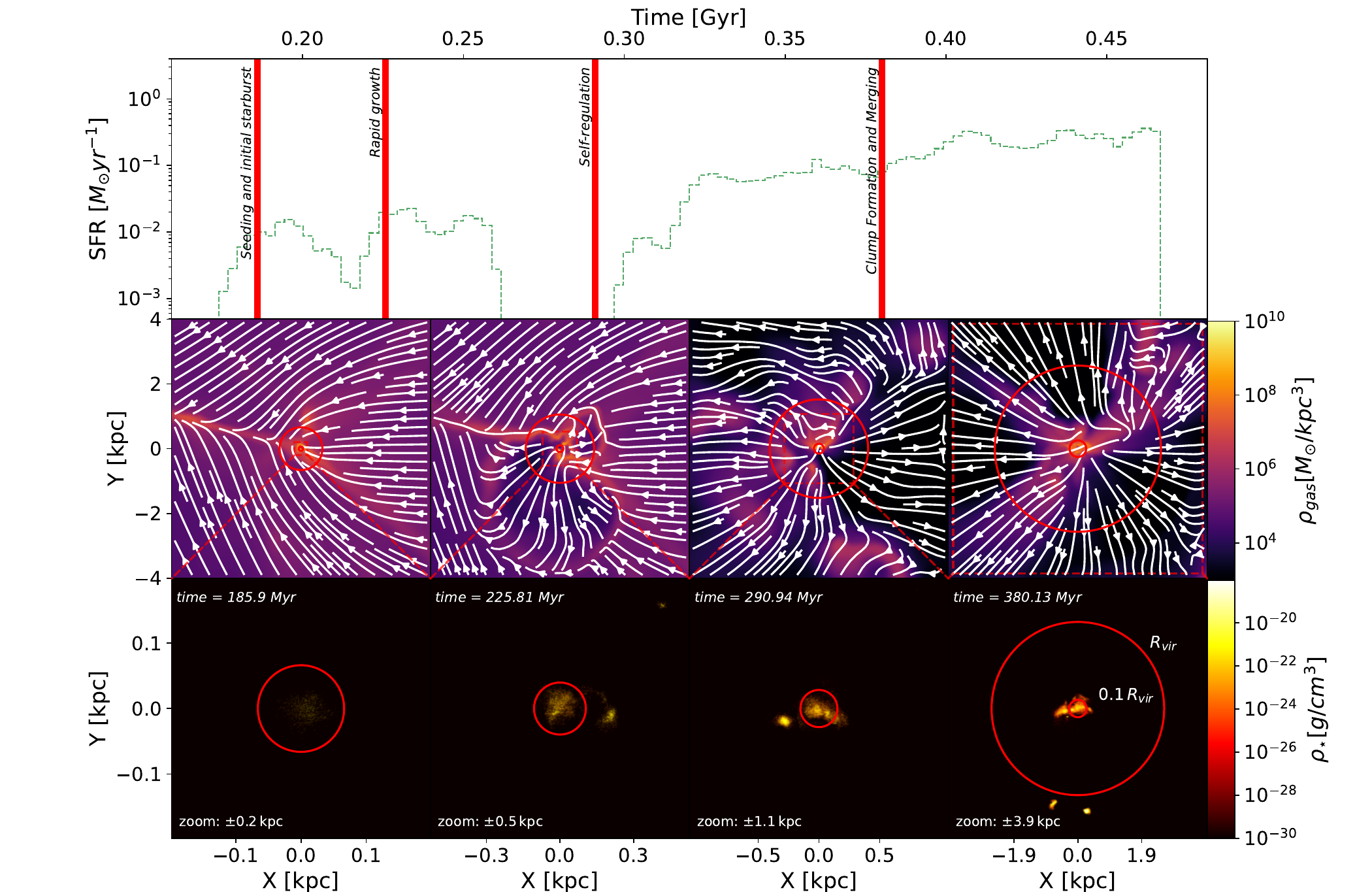} \\
    \caption{Star formation history (top) and snapshots from the AGN + kinetic SN feedback simulation with $E_{\rm SN} = 10^{51}\,\mathrm{erg}$. Vertical red lines mark four epochs ($t = 185.90$, $225.81$, $290.94$, and $380.13\,\mathrm{Myr}$) corresponding to distinct phases of BH–galaxy coevolution. Middle panels show $8\,\mathrm{kpc}$ slices of projected gas density with velocity streamlines. Bottom panels show zoomed-in stellar density maps of the central region. Dashed boxes indicate the zoomed regions, with connecting lines linking the panels. The zoom scale varies between snapshots (as labeled) to highlight central structure and, in the final panel, outer stellar clumps. Red circles denote $R_{\rm vir}$ and $0.1\,R_{\rm vir}$.}
    \label{fig:outflow}
\end{figure*}

\subsection{BH and Galaxy Formation at High Redshift}

In our previous work, we demonstrated that black hole (BH) growth at high redshift proceeds along a two-phase trajectory under super-Eddington accretion. The first phase is characterized by rapid, super-Eddington accretion, during which AGN feedback is not yet strong enough to halt gas inflows. As a result, the BH undergoes a period of accelerated mass growth. Once the BH mass reaches $\sim 10^{4}–10^{5}\,\msun$, feedback energy begins to dominate over radiative cooling, and the BH transitions into a self-regulated growth phase.

This two-phase picture can be compared to the recent scenario proposed by \citet{2026arXiv260114368P}, in which LRDs are interpreted as rapidly accreting black holes at high redshift. Both frameworks invoke an early phase of rapid black hole growth. The main difference lies in the modeling assumptions and focus: their work is designed to reproduce the observed properties of LRDs in a direct-collapse scenario, whereas in our simulations the early super-Eddington phase arises naturally from the gas supply and feedback conditions, and the model emphasizes the transition from this growth phase to subsequent self-regulation within a cosmological context.

The left column of Fig.~\ref{fig:data} illustrate these trends. Panel (b) shows BH mass as a function of time, while panel (c) presents the corresponding Eddington ratio, $\lambda_{\rm Edd} = \dot{M}_{\rm acc}/\dot{M}_{\rm Edd}$. Across all SN feedback models, the BH follows a broadly similar two-phase growth trajectory. In the high-energy SN case ($3\times10^{51}\,\mathrm{erg}$), early BH accretion is suppressed due to stronger SN-driven outflows, which delay gas cooling and reduce the supply of material to the BH. However, once cooling resumes, the BH growth converges to the same two-phase pattern observed in the lower-energy cases.

The right column of Fig.~\ref{fig:data} highlight the stellar component. Panel (d) shows stellar mass evolution, while panel (e) plots the star formation rate (SFR). Unlike BH growth, stellar mass assembly is highly sensitive to the SN feedback prescription. In the high-energy SN model (green solid line), the stellar mass at $z=10$ is nearly an order of magnitude lower compared to the lower-energy kinetic model (blue dotted line). The SFR also exhibits a more episodic and bursty pattern: stronger SN feedback extends the gas cooling timescale, resulting in longer quiescent intervals between bursts of star formation. This leads to substantially reduced stellar mass overall.

In contrast, AGN feedback has a more modest impact on star formation. As shown in panel (e), AGN activity primarily suppresses SFR near the transition between the rapid-growth and self-regulated phases, quenching star formation for $\sim 50\,\mathrm{Myr}$. Outside of this transitional period, AGN feedback has relatively little effect on the stellar mass history. Consequently, simulations with AGN feedback track the “no-feedback” case closely, while simulations with only SN feedback display markedly different stellar mass growth despite similar BH growth histories.

The results above can be understood within a simple physical picture of galaxy and BH co-evolution at high redshift. As halos grow, they accrete cold gas that fuels both star formation and BH accretion. At early times, when the gravitational potential is still shallow, SN feedback plays a dominant role in regulating baryonic processes. Stronger SN feedback increases gas cooling times, suppresses star formation, and may even prevent gas from collapsing efficiently into the central region. In extreme cases, such as low-mass halos experiencing strong supernova-driven outflows, gas fragmentation may delay or even prevent the formation of a central black hole seed.

Once gas can cool and accrete onto the black hole, the system enters a rapid super-Eddington growth phase. In this regime, supernova feedback primarily regulates star formation, while black hole accretion proceeds largely unaffected. When the black hole reaches $\sim 10^{4}$–$10^{5}\,M_\odot$, AGN feedback becomes strong enough to counteract cooling flows, marking the transition to the self-regulation phase. During this stage, AGN-driven outflows suppress further gas accretion onto the black hole and partially quench star formation, producing the valley-shaped star formation histories seen in our simulations. Over time, AGN-driven outflows persist, but star formation resumes in the outer halo, forming small stellar clumps that eventually merge into the main galaxy. Accretion of fresh gas via cold streams and galaxy mergers is frequent, so even when quenching occurs, star formation quickly recovers, and overall AGN feedback has limited effect on regulating star formation. Consequently, supernova feedback dominates the early baryonic environment and overall star formation efficiency, whereas AGN feedback primarily regulates the long-term balance of black hole growth.

To illustrate these processes, Fig.~\ref{fig:outflow} presents four snapshots of the kinetic SN feedback case with $E_{\rm SN}=10^{51}\,\mathrm{erg}$, taken at $t = 185.90$, $225.81$, $290.94$, and $380.13\,\mathrm{Myr}$. Each snapshot shows the gas density distributions and the stellar density, corresponding to distinct evolutionary stages identified in the SFR history (upper panel). In the first snapshot, the BH seed has just formed, coinciding with the first major burst of star formation. Gas continues to flow along filamentary structures toward the halo center. In the second stage, as the BH enters the super-Eddington phase, AGN feedback begins to compete with radiative cooling. Gas still accretes from the outer regions, but outflows become visible in the central region. By the third stage, the BH has transitioned into self-regulation: AGN feedback prevents gas inflows, temporarily quenching star formation for $\sim 50\,\mathrm{Myr}$. Finally, in the fourth snapshot, AGN-driven outflows persist, but star formation resumes in the outer halo, giving rise to small stellar clumps.

Taken together, these results underscore the complementary roles of SN and AGN feedback in shaping galaxy–BH co-evolution at high redshift. SN feedback dominates the assembly of stellar mass, introducing burstiness into the star formation history and regulates the early conditions for BH fueling, while AGN feedback governs the later self-regulated phase of BH growth.

\begin{figure*}[htbp]
    \center
    \includegraphics[width=\textwidth]{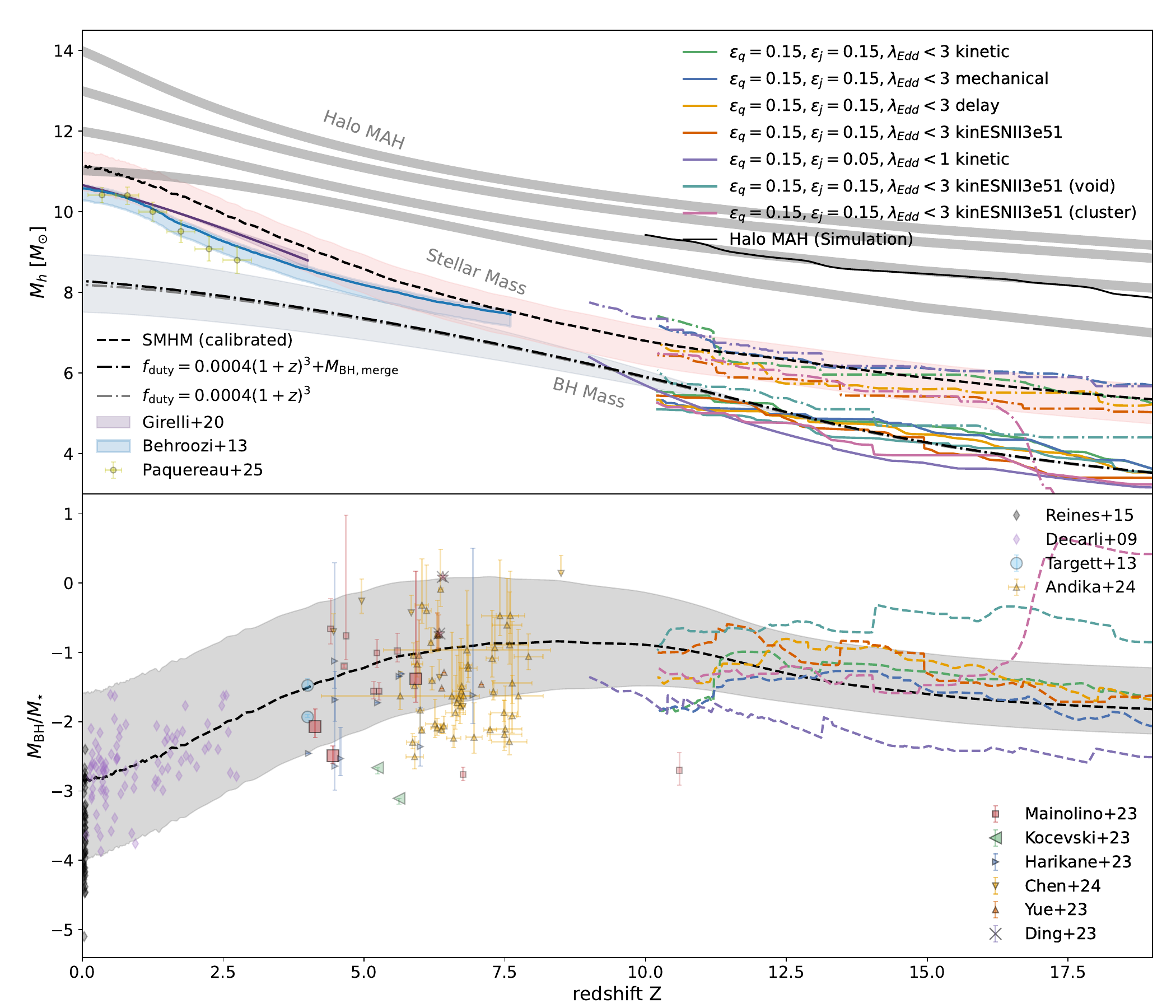}
    \caption{{\bf Top:}Overview of the self-consistent model. The halo assembly history from TNG is shown as the grey shaded curve, while the simulated halo mass evolution is plotted as the solid black line, reaching $M_{\rm halo} = 10^{12}\,M_\odot$ at $z = 0$. Analytical stellar mass growth using the TNG-calibrated SMHM relation is shown as a red shaded region, representing the uncertainty in stellar mass normalization arising from both observational systematics and variations among supernova feedback prescriptions. Stellar mass evolution from the individual feedback simulations is shown as colored dash–dotted lines. Independent stellar mass estimates from \citet{2020A&A...634A.135G} (purple), \citet{2013ApJ...770...57B} (blue), and \citet{2025A&A...702A.163P} (yellow circles) are also included. The black hole mass growth predicted by the observationally calibrated duty-cycle model is shown as the solid black curve, with the dark blue shaded region indicating the propagated $1\,\sigma$ uncertainty associated with the duty-cycle normalization. Contributions from black hole mergers are shown as a black dash–dotted curve, while black hole masses from the simulations are plotted as colored dash–dotted lines. {\bf Bottom:} Evolution of the $M_{\rm BH}/M_\star$ ratio for the feedback simulations (colored dashed curves) compared with the prediction from the self-consistent analytical model (black dashed curve). The grey shaded region denotes the propagated $1\,\sigma$ uncertainty in the mass ratio, combining uncertainties in both black hole growth and stellar mass normalization. From the peak to $z = 0$, the ratio decreases by roughly two orders of magnitude, exhibiting a broad peak at $z \sim 7$–10 and declining toward both higher and lower redshifts. The model also predicts that the mass ratio is higher than its $z=0$ value by factors of approximately (1.5, 3.0, 8.0, 35.5) at $z = (1, 2, 3, 5)$, respectively.}

    \label{fig:lowz}
\end{figure*}

\subsection{Pathways for the $M_{\mathrm{BH}}/M_{\star}$ Evolution}
To investigate the possible evolutionary pathways of the $M_{\mathrm{BH}}/M_{\star}$ ratio from $z=10$ to the present day, we develop a more refined and fully self-consistent analytical framework. This model links the rapid early growth of supermassive black holes (SMBHs) to the well-established scaling relations observed in the local universe. As discussed in Sec.~\ref{sec:overmass}, we construct an analytical model for the joint evolution of black hole mass and stellar mass. The BH growth model is calibrated directly against observational constraints, while the stellar mass evolution is tied to the halo mass assembly history and stellar-to-halo mass (SMHM) relation.

The star formation efficiency, $\epsilon_\star$, is calibrated using results from the IllustrisTNG simulations, or alternatively by matching the observed cosmic star formation rate density \citep{2014ARA&A..52..415M}. In this work, we adopt the TNG-based calibration, which naturally incorporates merger-driven stellar mass growth \citep{2019ComAC...6....2N}. Halo assembly histories are computed along the main progenitor branches of $z=0$ halos using EPS-based merger trees generated through Monte-Carlo sampling with Diffmah \citep{2021OJAp....4E...7H}, which has been tuned to reproduce the merger trees in TNG. 

For the black hole mass growth, in addition to the accretion described by our duty cycle model, we also include the contribution from black hole mergers. The total BH mass at time $t$ is therefore
\begin{equation}
M_{\rm BH}(t) = M_{\rm BH,0} + 
\int \dot{M}_{\rm BH}(t)\, dt
+ \sum M_{\rm BH,merge}(t),
\end{equation}
where the first term represents the seed BH mass, the final term represents the mass added through mergers with secondary black holes.

To estimate the merger-driven component, we assume that MBH mergers trace the merger history of their host halos. The halo merger rate is taken from the fitting formula of \citet{2010MNRAS.406.2267F}:
\begin{equation}
\frac{d N_m}{d \xi\, d z}(M, \xi, z)
= A\left(\frac{M}{10^{12} M_{\odot}}\right)^\alpha 
\xi^\beta 
\exp \left[\left(\frac{\xi}{\tilde{\xi}}\right)^\gamma\right]
(1+z)^\eta,
\end{equation}
where $M$ is the descendant halo mass, $\xi$ is the progenitor mass ratio, and the parameters $(A,\alpha,\beta,\gamma,\tilde{\xi},\eta)$ are taken from the best-fit values in \citet{2010MNRAS.406.2267F}. 

The corresponding BH mass growth rate from mergers is computed as
\begin{equation}
\dot{M}_{\rm BH,merge}(z) 
= 
\frac{
\int_{0}^{1} 
\frac{M_{\rm BH}(\xi M)}{t_{\rm df}}
\, dN_m(M,\xi,z)\, d\xi}{
\int_{0}^{1} dN_m(M,\xi,z)\, d\xi},
\end{equation}
where $t_{\rm df}$ is the dynamical friction timescale describing the orbital decay of the satellite galaxy and its central black hole within the host. Following \citet{2008ApJ...675.1095J}, we adopt
\begin{equation}
t_{\mathrm{df}}
=1.16\, 
\frac{M_{\mathrm{host}} / M_{\mathrm{sat}}}
{\ln \left(1+M_{\mathrm{host}} / M_{\mathrm{sat}}\right)}
f(\epsilon)\,
\left(\frac{r_c(E)}{r_{\mathrm{vir}}}\right)
t_{\rm{dyn}},
\end{equation}
with the orbital circularity dependence
\begin{equation}
f(\epsilon)=0.94\, \epsilon^{0.60}+0.60,
\end{equation}
and the dynamical time
\begin{equation}
t_{\rm dyn} \simeq 0.1\, H(z)^{-1}.
\end{equation}

Given the large uncertainties in modeling the detailed binary–hardening phase and the fact that the MBH coalescence time is generally shorter than the dynamical–friction timescale $t_{\rm df}$, we adopt the standard assumption that MBHs merge once their host galaxies undergo a major merger. This follows the common practice in semi-analytic models and captures the dominant contribution of major mergers to BH growth without introducing additional poorly constrained parameters \citep{2012MNRAS.423.2533B}. This choice is intended to provide a conservative upper limit on the contribution of mergers to black hole mass growth, as any realistic delays would further suppress the merger channel.
We note that this formulation also neglects the mass loss due to gravitational-wave emission during black hole coalescence, which is typically at the level of a few to $\sim 10\%$ \citep{2014PhRvD..90j4004H}. Given the relatively minor role of mergers in our model, this effect does not significantly impact our results.

In Fig.~\ref{fig:lowz}, we present an overview of our self-consistent model. The halo assembly history calibrated using the TNG simulations is shown as the grey shaded curve, while the halo mass evolution measured directly from our simulations is plotted as the solid black line. The latter closely follows the assembly history of a $10^{12}\,M_\odot$ halo at $z=0$, ensuring consistency between the analytic framework and the numerical simulations.

Using the TNG-calibrated stellar–halo mass relation (SMHM), we derive a redshift-dependent star formation efficiency, $\epsilon_{\star}(z)$. Our simulations employ four distinct supernova feedback prescriptions, which naturally produce a spread in stellar mass growth histories. To consistently relate this physically motivated spread to observationally inferred stellar masses, we introduce a uniform linear normalization factor $b_{\star}$. This factor is intended to capture systematic uncertainty in stellar mass normalization arising from both observational effects and feedback modeling, rather than to tune individual feedback prescriptions.

The star formation rate is therefore modeled as
\begin{equation}
\mathrm{SFR}(M_{\rm halo}, z) = b_{\star}\,\epsilon_{\star}(z)\, \frac{\Omega_b}{\Omega_m}\, \dot{M}_{\rm halo}(z).
\end{equation}

Instead of adopting a single normalization, we propagate stellar-mass uncertainty by allowing $b_{\star}$ to vary within a conservative range that brackets the lower and upper envelopes of stellar mass assembly produced by our different feedback simulations. This uncertainty is shown as the red shaded region in Fig.~\ref{fig:lowz}. Because $b_{\star}$ is applied uniformly in redshift, it affects only the overall normalization of the stellar mass and does not modify the relative evolutionary trends discussed below.

To assess the realism of this uncertainty range, we compare the resulting stellar masses with empirical determinations from \citet{2020A&A...634A.135G} (purple shaded region), \citet{2013ApJ...770...57B} (blue shaded region), and \citet{2025A&A...702A.163P} (yellow points). As shown in Fig.~\ref{fig:lowz}, observational estimates exhibit good agreement with simulation-based expectations at low redshift ($z \lesssim 1$) and substantially larger scatter at higher redshift. Importantly, all observational constraints lie within the red shaded envelope, supporting the interpretation that our feedback models and adopted normalization range plausibly capture the systematic uncertainty in stellar mass assembly.

For black hole growth, the BH mass evolution predicted by our observationally calibrated duty–cycle model is shown as the grey dash–dotted line. The BH masses from our feedback simulations are plotted as colored dash–dotted curves for comparison. We also show the predicted BH growth including the merger contribution, $M_{\rm BH,merger}$, as the black dash–dotted line. The duty–cycle model agrees well with our simulations, both exhibiting a two-phase evolutionary track characterized by an initial rapid, near-Eddington growth followed by a slower, self-regulated sub-Eddington phase. The simulated BH mass evolution closely follows the trajectory predicted by the duty–cycle model. In our framework, mergers contribute only $\lesssim 20\%$ of the total BH mass by $z=0$. Consistent with cosmological simulations, the contribution of BH–BH mergers to the overall SMBH mass budget is small: \citet{2015ApJ...799..178K} find integrated merger fractions of $\lesssim 3$–$23\%$, depending on environment, with SMBH growth predominantly driven by gas accretion, and the local SMBH mass function largely set by accretion processes \citep{2014MNRAS.440.1590D,2018MNRAS.476.2801M,2024ApJ...976....6Z}. Because our modeling assumptions effectively maximize the merger contribution, incorporating more realistic delays associated with dynamical friction, binary hardening, or wandering black holes would further reduce this fraction. A lower BH mass at low redshift would consequently decrease $M_{\rm BH}/M_{\star}$, thereby strengthening our conclusions. Even in the extreme case where black hole mergers are entirely suppressed following galaxy mergers, our main results remain unchanged.

In the lower panel of Fig.~\ref{fig:lowz}, we show the evolution of the $M_{\rm BH}/M_{\star}$ ratio for our feedback simulations (colored dashed curves) together with the prediction from our self-consistent analytical model (black dashed curve). The simulations closely follow the analytical expectation. The analytical model indicates that the local value of $M_{\rm BH}/M_{\star} \simeq 0.002$ at $z=0$ can be well reproduced. It also reveals a broad peak in the ratio at $z \sim 7$--$10$, reaching values from a few percent up to $\sim 30\%$, followed by a smooth, power-law–like decline toward $z = 0$.

We further compare our model with quasar observations spanning a wide range of redshifts, including data from \citet{2015ApJ...813...82R,2010MNRAS.402.2441D,2012MNRAS.420.3621T,2024A&A...685A..25A,2024A&A...691A.145M,2023ApJ...954L...4K,2023ApJ...959...39H,2025ApJ...983...60C,2024ApJ...966..176Y,2023Natur.621...51D}. The overall consistency between the observed black hole masses and our model predictions indicates that the inferred duty cycle is physically motivated rather than fine-tuned, and that it provides a natural evolutionary pathway for overmassive black holes at high redshift to evolve toward the local $M_{\rm BH}$–$M_\star$ relation.

Varying the duty-cycle normalization primarily shifts the overall normalization of black hole growth and the resulting $M_{\rm BH}$–$M_\star$ ratio. However, the qualitative evolutionary behavior is unchanged: across the full explored range, early black holes remain overmassive relative to their host galaxies at high redshift and subsequently evolve toward lower-redshift constraints following the same general trend.

Comparing the simulations in void and cluster environments with the fiducial case, we find that despite their different large-scale environments and assembly histories, both black hole growth and stellar mass growth exhibit qualitatively similar behavior at later times. At very high redshift, the void and cluster simulations show some variance relative to the fiducial run during the initial growth phase; however, the subsequent evolution follows comparable, well-defined tracks. As a result, the evolution of $M_{\rm BH}/M_\star$ is largely insensitive to environment within the halo mass range probed here.

When compared with the Eddington-limited simulation, we find that black hole growth in the Eddington-limited case remains in a rising phase down to $z \sim 9$,whereas in the duty-cycle model the effective accretion rate has already declined to $f_{\rm duty} \simeq 0.5$ by this epoch. This difference reflects distinct evolutionary trajectories in the two scenarios. Consequently, while the super-Eddington model already reproduces the transition from elevated $M_{\rm BH}/M_\star$ ratios at high redshift to lower values at later times, our current simulation window does not allow us to determine whether the Eddington-limited case would undergo a similar self-regulation phase at lower redshift. Importantly, however, both accretion prescriptions yield qualitatively similar trends at high redshift, demonstrating that the presence of elevated $M_{\rm BH}/M_\star$ ratios does not depend on finely tuned accretion assumptions.

Taken together, these comparisons indicate that a scenario characterized by rapid early growth followed by self-regulated accretion naturally explains the observed transition from elevated $M_{\rm BH}/M_\star$ ratios at high redshift to lower values at later times, while remaining consistent with current observational constraints.

Overall, our self-consistent model provides a physically grounded framework that links the rapid early growth of SMBHs at $z > 10$ to the much lower $M_{\rm BH}/M_\star$ ratios observed today. It shows how systems hosting apparently overmassive black holes at high redshift can naturally evolve onto the canonical local scaling relation by $z = 0$.

\section{Conclusion} \label{sec:conclusion}
We have investigated the co-evolution of black holes and their host galaxies in the high-redshift universe using cosmological hydrodynamic simulations with varying supernova (SN) feedback prescriptions. Our results show that black hole growth is largely insensitive to the strength of SN feedback, whereas the stellar mass of the host galaxy is strongly regulated. In particular, adopting a higher single SN energy ($3\times 10^{51}\,\mathrm{erg}$) suppresses star formation more effectively, producing galaxies at $z=10$ with stellar masses an order of magnitude lower than in weaker-feedback models, while central black holes reach similar masses. This differential response naturally explains the elevated black hole–to–stellar mass ratios, $M_{\rm BH}/M_\star$, inferred from JWST observations. Strong SN feedback at high redshift—arising from energetic explosions or the shallow gravitational potentials of early galaxies—drives outflows that suppress star formation without significantly hindering black hole growth, providing a plausible pathway to reconcile overmassive SMBHs with galaxy–SMBH co-evolution models.

However, the impact of SN feedback on star formation in high-redshift galaxies remains an active area of investigation. For example, \citet{2014PhRvD..90j4004H} emphasize the role of stochastic, bursty star formation in low-mass halos, which can enhance the observed UV luminosities through short-lived star formation episodes and thereby reduce the apparent effectiveness of feedback in suppressing star formation. While such burstiness can temporarily boost the UV output, our results show that strong SN feedback still suppresses the overall build-up of stellar mass. These effects are therefore complementary, with bursty star formation affecting instantaneous UV luminosities, while SN feedback regulates the long-term growth of stellar mass.

At the high redshift, we identify four characteristic stages of BH–galaxy co-evolution: (i) {\bf Seeding and Initial Starburst}, where the BH seed forms alongside the first major star formation burst; (ii) {\bf Rapid Growth}, characterized by a super-Eddington BH phase while SN feedback regulates star formation; (iii) {\bf Self-Regulation}, when BHs reach $\sim 10^{4}$–$10^{5}\,\msun$ and AGN feedback balances inflows, quenching star formation for $\sim 50\,\mathrm{Myr}$; and (iv) {\bf Clump Formation and Merging}, during which AGN outflows continue, but star formation resumes in the outer halo, forming stellar clumps that eventually merge into the central galaxy.

To place this high-z finding in a broader cosmological context,
we developed an analytic model for SMBH and stellar mass growth at lower redshifts,
where stellar mass follows the halo assembly history, while BH mass is modeled using a duty cycle calibrated by observational AGN samples at $z=0-6$, with an effective Eddington-rate duty cycle $f_{\rm duty} = 0.0004 (1+z)^3$. This prescription yields a two-phase BH growth trajectory seen in simulations: an early rapid, near-Eddington phase followed by a slower, sub-Eddington self-regulated phase. The model successfully explains the overmassive BHs seen in our simulations.

We further calibrate our model using the SMBH–stellar mass relation from IllustrisTNG and include BH merger contributions, which allow the relation to evolve to low redshift ($z=0$). The resulting stellar mass trajectories agree with independent low-redshift estimates, and the predicted $M_{\rm BH}/M_\star$ ratio reproduces the local relation, $M_{\rm BH}/M_\star \simeq 0.002$ at $z=0$. Our model predicts $M_{\mathrm{BH}}/M_{\star}=(0.002,0.003,0.006,0.016,0.071,0.156)$ at 
$z=(0,1,2,3,5,10)$. These values are broadly consistent with existing observational constraints, including the elevated ratios inferred in the JWST era. This redshift evolution naturally emerges from the combined effects of early rapid BH growth, declining duty cycles at later times, and the continued buildup of stellar mass after the peak of BH accretion activity. As a result, the model captures both the high $M_{\rm BH}/M_\star$ ratios observed in high redshift galaxies and the gradual convergence toward the tight local relation at $z=0$. The consistency of the results across halos in filament, void, and cluster environments suggests that the proposed growth pathway is not strongly dependent on large-scale environment at fixed halo mass.

While our fiducial model adopts a super-Eddington accretion prescription, Eddington-limited growth represents an equally plausible physical limit at high redshift. Rather than attempting to distinguish between these modes, we use them to bracket the range of possible black hole growth histories.

Within the redshift range directly probed by our simulations ($z \gtrsim 9$), the two scenarios correspond to different evolutionary phases. In the Eddington-limited case, the black hole mass growth is still increasing at $z = 9$ and shows no evidence of self-regulation. In contrast, the super-Eddington model undergoes rapid early growth followed by a self-regulated phase, leading to a declining $M_{\rm BH}/M_\star$ ratio toward lower redshift.

Although our extrapolated duty-cycle model predicts a lower $M_{\rm BH}/M_\star$ ratio by $z \sim 10$ than implied by continued Eddington-limited growth, we emphasize that distinguishing between these scenarios would require simulations extending to lower redshift. Consequently, our results do not rely on identifying the correct accretion mode, but instead demonstrate that a self-regulated growth phase naturally reproduces the observed transition from elevated $M_{\rm BH}/M_\star$ ratios at high redshift to lower values at late times.

Independent of the accretion prescription, we robustly find that $M_{\rm BH}/M_\star$ exhibits a broad peak at $z \sim 7$–10, with values ranging from a few percent up to $\sim 30\%$, followed by a steady, approximately power-law decline toward $z = 0$.

Our study thus demonstrates that a simple, observationally motivated duty-cycle prescription, when extrapolated in a conservative and explicitly uncertain manner, can plausibly connect early SMBH growth to lower-redshift constraints. Super-Eddington accretion at early times, combined with a declining duty cycle at later times, naturally produces qualitative trends consistent with both high-redshift observations and the local SMBH–stellar mass relation. While the precise normalization of the extrapolated model remains uncertain, the robustness of our conclusions lies in these qualitative evolutionary trends. Future work including more detailed modeling of AGN feedback and a larger sample of simulated halos will help further test this pathway. Testing this picture against additional observational data will also strengthen these conclusions.

\section{Acknowledgments}
\begin{acknowledgments}
We acknowledge the support from the National Key Research and Development Program of China (No.2022YFA1602903), the NSFC \ (No. 11825303, 12347103, 11861131006), the science research grants from the China Manned Space project with No. CMS-CSST-2021-A03, CMS-CSST-2021-B01, the Fundamental Research Funds for the Central Universities of China \ (226-2022-00216), the start-up funding of Zhejiang University and Zhejiang provincial top level research support program. R.C. and Z.W. thanks Princeton University for hospitality during a visit when this work was initiated.
We also acknowledge the cosmology simulation database (CSD) in the National Basic Science Data Center (NBSDC) and its funds the NBSDC-DB-10. The simulations and analysis presented in this article were carried out on the SilkRiver Supercomputer of Zhejiang University. 
\end{acknowledgments}

\bibliography{reference}{}
\bibliographystyle{aasjournal}



\end{document}